\documentclass[12pt]{article}
\usepackage{epsfig}
\usepackage{graphicx,color}
\usepackage[normalem]{ulem}

\newcommand{\beq}{\begin{equation}}
\newcommand{\eeq}{\end{equation}}
\newcommand{\beqa}{\begin{eqnarray}}
\newcommand{\eeqa}{\end{eqnarray}}
\newcommand{\bd}[1]{ \mbox{\boldmath $#1$}}

\begin{document}
\def\ii{\'\i}

\title{
The pseudo-Semimicroscopic Algebraic Cluster Model model: Heavy nuclei
}

\author{H. Y\'epez-Mart\ii nez$^1$, P. O. Hess$^2$ \\
{\small\it
$^1$Universidad Aut\'onoma de la Ciudad de M\'exico,
Prolongaci\'on San Isidro 151,} \\
{\small\it
Col. San Lorenzo Tezonco, Del. Iztapalapa,
09790 M\'exico D.F., Mexico} \\
{\small\it
$^2$Instituto de Ciencias Nucleares, UNAM, Circuito Exterior, C.U.,}\\ 
{\small\it A.P. 70-543, 04510 M\'exico D.F., Mexico} }

\maketitle
\abstract{
The Semimicroscopic Algebraic Cluster Model (SACM) is extended to heavy
nuclei, making use of the pseudo-$SU(3)$ model. As a first step, the concept of 
{\it forbiddenness} will be resumed.
One consequence of the {\it forbiddenness} is that the
ground state of a nucleus can in general be described by two internally excited clusters.
After that, the pseudo-SACM is formulated. The basis of pseudo-SACM
is constructed, defining each cluster within the
united nucleus with the same oscillator frequency and deformation of the harmonic oscillator as a mean field
and dividing the nucleons in those within the unique and 
normal orbitals, consistently for both clusters and the united nucleus. 
As test cases, this model is applied to 
$^{236}$U $ \rightarrow$ $^{210}$Pb+$^{26}$Ne and
$^{224}$Ra $\rightarrow$ $^{210}$Pb+$^{14}$C. Some spectroscopic
factors will be calculated as predictions.
}

\vskip 1cm

\section{Introduction}
\label{intro}

The {\it Semimicroscopic Algebraic Cluster Model} (SACM) was introduced in
\cite{cseh-letter, cseh-levai-anph} and applied to light nuclei up to the
first half in the sd-shell.  Later some attempts were made to extend it to heavy nuclei.
Many applications of the SACM have been studied since then, for example, a construction of the effective $SU(3)$ irreducible
representations (irreps)  for heavy nuclei \cite{hunyadi}, using the 
Nilsson model, and a study of preferences in radioactive decays and/or fission
\cite{sacm-fission1,sacm-fission2,sacm-fission3}.   
More recently \cite{hess-86}, with the help of the SACM the 
spectra of
$\alpha$ cluster nuclei, of great interest in astrophysics related to the production  
of heavy elements, and their spectroscopic factors were calculated. In \cite{phase-I,phase-II}
the phase transition properties of the model were investigated and in \cite{renorm}
the renormalization of the coherent state parameters, used for the geometric
mapping, was presented.

Though, in \cite{hunyadi} a quite powerful method was presented on how to treat heavy
nuclei, it only delivers the {\it ground state}, or the first super- and hyper-deformed states
\cite{cseh2006}. Therefore, it is of interest to
look for different procedures  to deal also with excited states. 
One of those is the {\it pseudo-$SU(3)$ model}
($\widetilde{SU}(3)$) \cite{hecht,arima}.
In \cite{cseh-scheid,cseh-algora} the first attempts have been made to extend
the SACM to heavy nuclei,  %using just %this pseudo-$SU(3)$ (
applying this $\widetilde{SU}(3)$ model.  
Problems arise with not reaching the ground state of the united nucleus, after having
coupled the two clusters and the relative motion,
which turned out to be forbidden in the approximation of the leading representation. 
In standard $SU(3)$, for clusters with approximately more than 12 protons or neutrons, there is
in general
no overlap between the coupled cluster state and the relative motion with 
the ground state of the united nucleus. This was first
pointed out in \cite{smirnov}, where the notion of {\it forbiddenness} was introduced.
There, the {\it forbiddenness} is defined as the minimal number of excitation quanta needed 
in at least one of the
clusters, such that an overlap with the united  nucleus is possible. 
Reconsidering the definition of {\it forbiddenness} from an alternative angle, we were able to prove 
\cite{huitz-2015} that the numbers determined in \cite{smirnov} contained
numerical errors. In this contribution we will resume the result briefly.

In  \cite{cseh-scheid,cseh-algora} a different but equivalent definition to the {\it forbiddenness} 
was given, which will be introduced further below.
Also in \cite{cseh-scheid,cseh-algora}, 
the separation of nucleons into the unique and normal orbitals in the united 
nucleus, compared to the ones in each cluster, is not well defined:
Distinction is made when one cluster is light, which is then treated within the
standard shell model, or both are heavy.

Both models, the $\widetilde{SU}(3)$ and the pseudo-SACM, will be explained 
briefly in section \ref{pseudo-sacm} and   the problems related to the direct extension 
of the pseudo-$SU(3)$ scheme to the SACM are mentioned.

After having introduced the improved determination of 
the {\it forbiddenness}, the properties of the $\widetilde{SU}(3)$ model and the SACM,  for light nuclei
and in its version for heavy nuclei in 
\cite{cseh-scheid,cseh-algora}, will be reviewed. 
We will proceed in section \ref{fil} presenting a possible alternative, namely the
{\it pseudo-SACM}. 
In order to show the utility of the pseudo-SACM, in section 
\ref{examples} the new proposal is applied to two systems, namely 
$^{236}$U $ \rightarrow$ $^{210}$Pb+$^{26}$Ne and
$^{224}$Ra $\rightarrow$ $^{210}$Pb+$^{14}$C.
 In section \ref{conclusions} conclusions are drawn.

\section{The pseudo-SACM}
\label{pseudo-sacm}

The pseudo-$SU(3)$ model \cite{hecht,arima} 
is based on the  near degeneracy observed in the Nilsson scheme, when in
each harmonic oscillator shell $\eta$
the orbital belonging to the
largest spin $j=\eta + \frac{1}{2}$ is skipped from
consideration.
To the remaining orbitals the redefinition

\beqa
j ~=~ l\pm\frac{1}{2} & \rightarrow & {\tilde l}~=~ 
l\mp\frac{1}{2}
\nonumber \\
\eta & \rightarrow & {\tilde \eta} ~=~ \eta - 1
~~~,
\label{eq-1}
\eeqa
is applied, where ${\tilde l}$ denotes the
{\it pseudo-orbital angular momentum}
and ${\tilde \eta}$ the pseudo-shell number.
Those orbitals with {\it the same pseudo-orbital angular momentum} ${\tilde l}$ are 
nearly degenerate,
which implies a very small pseudo-spin-orbit interaction. In addition, the content
of the $\widetilde{\eta}$ shell corresponds to the one of $\eta = \widetilde{\eta}$ in the
standard shell model.
Thus, the shell model for light
nuclei can be directly extended to heavy nuclei, using $\widetilde{SU}(3)$ model instead of $SU(3)$ model. The orbitals renamed by the pseudo-orbital spin are called {\it normal orbitals},
while those in the orbitals with maximal spin $j=\eta+\frac{1}{2}$ are called
{\it unique  or intruder levels}. Nucleons in these unique orbitals are treated as {\it spectators},  i.e., it
is supposed that the particles in the intruder levels follow
the dynamics of those in the normal orbitals and
thus they are not forgotten but rather taken into account through well defined effective charges
\cite{pseudo-sympl}.
The influence of the nucleons in the intruder levels are also indirectly taken into account
via the parameters of the model. As an example, the intruders are important to obtain correct
collective masses, but because these masses are parameters within the model, the effect is taken
into account implicitly. Other effects, as the back-bending mechanism
provoked by the decoupling of a fermion pair in the intruder orbitals, are not included in the model. 

In \cite{pseudo-sympl} the basic assumptions of  the $\widetilde{SU}(3)$ model  were  extended to
the {\it pseudo-symplectic} model of nuclei, which takes into account the nucleons
in the closed shells via inter-shell excitations, while the nucleons in the unique
orbitals are treated still as spectators. The effective charges are not
parameters, as in the $SU(3)$ model, but 
represent scaling factors, which 
has a definite dependence on the total number of nucleons
and on the division of these nucleons into those in the unique and normal orbitals. There is a similarity between the SACM and the symplectic model \cite{rowe1}, namely that both include
excitations of $2\hbar\omega$ in the Hamiltonian.

In the SACM for light nuclei, first the $SU(3)$ irrep are determined in the usual way, i.e.
each cluster is represented by an irrep $\left( \lambda_k,\mu_k\right)$ ($k=1,2$)
for a two cluster system in their ground state. 
Adding the number of oscillation quanta in each cluster and
comparing them with the number of oscillation quanta of the united nucleus results in
a mismatch: The number of oscillation quanta of the united nucleus is larger than
the sum of both clusters. Wildermuth \cite{wildermuth} showed that the necessary condition
to satisfy the Pauli exclusion principle is to add 
the missing quanta into the relative motion, 
introducing a minimal number of relative 
oscillation quanta $n_0$.
This is known as the {\it Wildermuth condition}.  However, there are 
still irreps which are not allowed by the Pauli-exclusion principle. 

An elegant solution to it, avoiding cumbersome {\it explicit} antisymmetrization of the wave function, was proposed \cite{cseh-letter,cseh-levai-anph}, where the SACM was presented for the first time: 
The coupling of the cluster irreps with the one of the relative motion generates a list of $SU(3)$ irreps, i.e.,

\beqa
\left(\lambda_1,\mu_1\right) \otimes \left(\lambda_2,\mu_2\right) \otimes
\left(n_\pi , 0\right) & = & \sum_{m_{\lambda , \mu} } m_{\lambda , \mu} 
\left( \lambda , \mu \right)
~~~,
\label{eq-2}
\eeqa
where $n_\pi$ is the number of relative oscillation quanta, limited from below
by $n_0$ and $m_{\lambda , \mu}$ is the multiplicity of
$\left( \lambda , \mu \right)$.

This list of irreps is compared to the one of the shell model. Only those
are included in the SACM model space, which have a counterpart in the shell model.
In such a way, the Pauli exclusion principle is observed and the model space can be  called  microscopic.

The word {\it Semi}  in the name of SACM appears due to the phenomenological character
of the Hamiltonian, which is a sum of terms related to single particle energies, 
quadrupole-quadrupole interactions, angular momentum operators and more. In section \ref{fil}
the structure of the Hamiltonian
will be exposed and explained.

As already mentioned, a first attempt to extend
the SACM to heavy nuclei was published 
in \cite{cseh-scheid,cseh-algora}.
The procedure is very similar to the one for light nuclei, 
with the difference that now only nucleons in
the normal orbitals are considered. 
For the case of two heavy clusters, for each cluster
the nucleons are filled into the Nilsson scheme
at the deformation of the corresponding cluster and the same is done for the  parent 
nucleus. When a unique orbital is filled, the
nucleons are excluded from counting, while when a normal orbital is filled they are included.
As an alternative to this procedure, we use the deformation of the united nucleus also for the clusters
(see further detailed discussion in Section \ref{fil}).
In this manner, filling in the protons and neutrons, 
each cluster has a given number of nucleons in the normal and
in the unique orbitals. The $\widetilde{SU}(3)$ irreps for each cluster are determined, using only the nucleons in the normal orbitals, i.e., restricting to the pseudo-oscillator. These irreps
are coupled with each other and the one of the relative oscillator, yielding a list of final irreps similar to (\ref{eq-2}). The problem with 
this procedure is that {\it even for one light cluster} the obtained list may have 
{\it no overlap}\ with the ones
of the irreps in the united nucleus, though some come closer than others. 

In order to be able to deal heavy systems, an alternative definition of the forbiddenness
was proposed in \cite{cseh-scheid,cseh-algora}, where the relation of an irrep to its deformation 
was exploited.
Thus, if an irrep of the list has a similar deformation as the one in the shell model allowed irrep, one can say that it is {\it less forbidden} than 
irreps with a larger difference in the irreps. Therefore, the notion of forbiddenness used 
in \cite{cseh-scheid,cseh-algora} is 

\beqa
F & = & \frac{1}{1+min\left[\sqrt{\Delta n_1^2+\Delta n_2^2+\Delta n_3^2}\right]}
~~~,
\label{eq-3}
\eeqa
where $\Delta n_i = \mid n_i - n_{i,k}\mid$
and in contrast to $S$, as defined in  
\cite{cseh-scheid,cseh-algora}, we use $F$ because the letter
$S$ will be used later on for the spectroscopic factor. 
The index $i$ refers to the 
spatial direction of the oscillation and $k$ to the  several cluster irreps allowed by  the Pauli-exclusion principle in (\ref{eq-2}). The $n_i$ is the
number of oscillation quanta in direction $i$. This is a distinct definition
of {\it forbiddenness} as in \cite{smirnov}.

It is important to mention that for the case when the parent nucleus consists of one heavy 
and a light  cluster, the light cluster is treated up to now  
within the $SU(3)$-model, i.e., 
a mixture of models ($SU(3)$ and $\widetilde{SU}(3)$)
is used,
adding to the arbitrariness of dividing
the nucleons into those occupying unique or normal orbitals. Also, in general,
the number of nucleons in the normal orbitals of the clusters do not match
those in the united nucleus. 
Nevertheless, 
this concept proved to be quite  useful  in the understanding of structural preferences 
in the fission or radioactive decay 
\cite{sacm-fission1,sacm-fission2,appl-fission1}.

In spite of this success, it is to us not very satisfactory to 
deviate in such an amount from the
original version of the SACM, where the Pauli principle was taken into account within 
the {\it same harmonic oscillator mean field}.
This is the reason why we started to reanalyze the SACM in its
version for heavy nuclei, where the spin-orbit interaction is taken into account
effectively within the $\widetilde {SU}(3)$ model.

\subsection{A new concept of forbiddenness}
\label{forbiddeness}

In \cite{smirnov} it was shown that putting all 
missing oscillation quanta only into the relative motion,
with increasing mass number of the lightest cluster, 
from one point on it will
not be possible to couple
the two clusters with the relative motion to the irrep of
the united nucleus, thus, it is {\it forbidden}.
This property is due to the too large irrep $(n_\pi ,0)$
of the relative motion and $n_\pi >n_0$ increases
quickly with a larger cluster mass.

In heavy nuclei valence protons and neutrons occupy different shells and 
one has to treat them separately. 
First, we will resume the equations to solve.
One way to solve the problem is to allow
excitations of one  or two of the clusters and adding
the remaining oscillation quanta to the relative 
motion \cite{cseh-letter}. 
The minimum number of oscillation quanta needed, to achieve 
a final overlap with the ground state irrep of the united nucleus,
is called the {\it forbiddenness}.
Unfortunately, it turned out to be difficult to follow the arguments given in \cite{smirnov} on how to
determine the {\it forbiddenness}. For this reason, the authors
published in \cite{huitz-2015} a detailed description 
for the 
determination of the {\it forbiddenness} and we resume
the main result only (in heavy nuclei, the quantum number in the equation have to
carry an addition index for protons or neutrons):

\beqa
n_{C} & = & 
{\rm max}\left[0,\frac{1}{3}\left\{ n_{0}-(\lambda -\mu
)-(2\lambda _{C}+\mu _{C})\right\} \right] 
\nonumber \\
&& +{\rm max}\left[ 0, \frac{1}{3}\left\{ n_{0}-(\lambda +2\mu )
+(\lambda _{C}-\mu_{C})\right\} \right]
~~~.
\label{forbid-min}
\eeqa
In (\ref{forbid-min}) the $n_C$ denotes the {\it forbidenness},
$(\lambda_C,\mu_C)$ the cluster irrep (to which the two clusters
are coupled and there may appear several), $n_0$ is the {\it Wildermuth condition} and
$(\lambda , \mu )$ is the final $SU(3)$ irrep of the
united nucleus. The total number of quanta $n_0$, according to the {\it Wildermuth condition},
is the sum of $n_C$ and the remaining relative oscillation quanta.
For later use, we define $\left(\lambda_0, \mu_0\right)$ as the difference of the final cluster
irrep $\left(\lambda^e_C,\mu^e_C\right)$ to the former one 
$\left(\lambda_C,\mu_C\right)$ (the letter $e$ stands for {\it excited}), via

\beqa
\left(\lambda^e_C,\mu^e_C\right) & = & \left(\lambda_C + \lambda_0,\mu_C+\mu_0\right)
~~~. 
\label{la0mu0}
\eeqa

Eq. (\ref{forbid-min}) 
can be interpreted as follows: 
The first term in (\ref{forbid-min}) tells us, that 
in order to {\it minimize} $n_C$, we
have to {\it maximize} $(\lambda_C+2\mu_C)$.
The second term tells us that in addition the difference
$\left(\lambda_C - \mu_C\right)$ has to be minimized. The condition of a maximal 
$(\lambda_C+2\mu_C)$ and a minimal $\left(\lambda_C - \mu_C\right)$
implies a large compact and oblate 
configuration of the two-cluster system.

One can achieve these conditions, 
determining the whole product
of $(\lambda_1,\mu_1) \otimes (\lambda_2,\mu_2)$ 
and searching for the irrep that corresponds to a large
compact structure (large $\left(2\lambda_C+\mu_C\right)$ but with a maximal 
difference $\left(\mu_C-\lambda_C\right)$).
For deformed clusters, there is also 
the possibility to excite it within the $0\hbar \omega$, leading to other 
individual cluster irreps $(\lambda_C,\mu_C)$. 
One can take the whole 0$\hbar\omega$ space of
each cluster and multiply them all, or even one can do it for the proton space and
the neutron space for each cluster and then multiply the proton final space with
the neutron final space, which has to be done anyhow for heavy nuclei because protons
and neutrons are in different shells. 

The result (\ref{forbid-min}) is similar in structure for light as for
heavy nuclei, changing the $SU(3)$ irreps
by their $\widetilde{SU}(3)$ values.
Before continuing, we shall discuss the philosophy of
the extension of the SACM to the pseudo-SACM model, which
addresses the question on how to define a cluster 
within the pseudo-$\widetilde{SU}(3)$ shell model.

We checked the relation of the definition in (\ref{eq-3}) to the above definition
of the forbiddenness and the qualitative consequences {\it are the same}, i.e.,
what is forbidden in one definition is also forbidden in the other one and the same for the
allowed irreps. The difference is the explicit determination of $n_C$, using (\ref{forbid-min}).

\section{Some basic philosophical changes}
\label{fil}

In this section, we propose an alternative 
semimicroscopic description of cluster states in heavy nuclei,
which is consistent in the separation of nucleons occupying the normal and unique orbitals,
for the two clusters as for the united nucleus.

The main point of the proposal, which we like to stress, 
is that when the united nucleus is considered as a sum
of two clusters, these clusters are defined {\it within the united nucleus} and
{\it not as individual clusters}. This happens already in the SACM for light nuclei: The
fundamental scale of the harmonic oscillator of the shell model, $\hbar\omega$, is
{\it the same} for each cluster and the relative oscillator. For example, in
$^{16}$O+$\alpha$ $\rightarrow$ $^{20}$Ne, the $\hbar\omega$ used is 
$45 A^{-1/3}-25 A^{-2/3}$ = 13.19~MeV  \cite{hw} 
for the united nucleus $^{20}$Ne, while the value for
the free $^{16}$O is  13.92 ~MeV and for $\alpha$ it is 18.43~MeV. The differences are quite large!
The argument why the $\hbar\omega$ value of the united nucleus has to be taken is that
the two clusters happen to be formed within the {\it united} nucleus, i.e, the same mean field.

In \cite{cseh-scheid,cseh-algora}   the present approach  was followed only partially, certainly not
for the criteria of which nucleons are in the normal or unique levels. The light clusters were treated within the
$SU(3)$, while the heavy ones are treated within
$\widetilde{SU}(3)$ model.
This leads to an inconsistent separation of  active and non-active nucleons. 

We propose that a consistent way is to continue to treat each cluster as an entity 
{\it within} the united nucleus, which means:

\begin{itemize}

\item In order to determine the number of nucleons in the normal orbitals,
the Nilsson level scheme is used and the orbitals are filled at 
a {\it fixed} deformation. We propose to use {\it the same deformation} 
of the united nucleus also for the clusters. 

\item The nucleons of the heavy cluster are filled in first.
This gives the number of nucleons
in normal orbitals for this first cluster. Then, {\it on top of it},  
the nucleons of the light cluster 
are filled into
the Nilsson scheme until the united nucleus is reached. 
The number of nucleons in the
normal orbitals of this united nucleus minus the number of nucleons in normal orbitals of
the heavy cluster gives as a result the number of normal nucleons of the light cluster.
This assures that the so-called active nucleons in the united nucleus are equal to the
sum of active nucleons of the two clusters.
Considering that we are interested either in the {\it pre-formation} of a light cluster
or in the collision of a light cluster as a projectile, the light cluster is considered
always as being added on top of the heavy cluster. This will be important when we
construct the model space. 

\item Once the number of nucleons in the normal orbital for each cluster (${\tilde A}_1$ and ${\tilde A}_2$)
is obtained, these are
filled into the pseudo-shell model.

\item The {\it Wildermuth condition} is applied to the  pseu\-do-oscilla\-tor,   i.e., the minimal
number of oscillation quanta needed is the difference of the oscillation quanta in the
pseudo-oscillator  of the united nucleus to the sum of oscillation quanta of the
 pseudo-oscillator  of the two clusters.
Analogous as Wildermuth showed \cite{wildermuth}, the pseudo-shell model of the united nucleus is
related  by an orthogonal transformation
to the two pseudo-shell models of the clusters plus the oscillator for the relative motion.
Thus, from a mathematical point of view there
is no ambiguity. 
In fact one can proceed now in the same way as in the SACM
for light nuclei, due to the assured match of the overlap 
of the irreps of the two clusters (though excited) 
with the relative motion to the leading representation of the parent nucleus.

\item States in the united nucleus are described by the product of two clusters,
which in general are excited,
and the number of quanta which remain in the relative motion.

\item The nucleons in the unique orbitals are not forgotten, rather their dynamics are taken into account through a
scale factor which can be interpreted as an effective charge
and also indirectly through the parameters of the model.

\end{itemize}

The advantage of this procedure is obvious: The Pauli exclusion principle is maintained and
the elegance of the SACM for light nuclei is transferred to heavy nuclei. No practical
problems appear, no mixing of different oscillator models is needed 
and the interpretation remains clear.

\subsection{The structure of the Hamiltonian}
\label{hamiltonian}

Concerning the construction of the model space, 
equation (\ref{eq-2}), for light clusters, changes to

\beqa
&
\left({\tilde \lambda}_C,{\tilde \mu}_C\right) \rightarrow
\left({\tilde \lambda}^e_C,{\tilde \mu}^e_C\right)
&
\nonumber \\
&
\left({\tilde \lambda}^e_C,{\tilde \mu}^e_C\right) 
\otimes
\left({\tilde n}_r , 0\right)  =  
\sum_{m_{{\tilde \lambda} , {\tilde \mu} }} m_{{\tilde \lambda} , {\tilde \mu}} 
\left( {\tilde \lambda} , {\tilde \mu} \right)
~~~,
&
\label{eq-4}
\eeqa
for heavy clusters,
where $\left({\tilde \lambda}^e_C,{\tilde \mu}^e_C\right)$ refers to the excited 
cluster irrep.
No irreps for the individual clusters are mentioned yet, because this is a more complicated
matter, involving the cluster irreps for protons and neutron separately, which are
afterward coupled to a cluster irrep 
$\left( \tilde{\lambda}_C,\tilde{\mu_C}\right)$
of the combined system. 
The index $e$ refers to the excited cluster irrep, as explained in the subsection 
2.1 on {\it forbiddenness}.
The list of the irreps of the $\widetilde{SU}(3)$  model  are compared to the shell model irreps
of the pseudo-oscillator.   Only those irreps which appear in the
pseudo-oscillator are included in the $\widetilde{SU}(3)$ model space. However, for heavy systems
the model space is still extremely large and one has to apply further simplifications. Because
this is a matter for itself, the details will be explained in subsection \ref{sub-3.3}.
There, we will propose further restrictions on how to reduce the size of the model space,
using physical arguments.

The most general algebraic Hamiltonian has the  same structure as for light nuclei, except that the
operators (number operator, quadrupole operator, etc.) are substituted by their pseudo-counter
parts. How this is done, is explained in detail in 
\cite{draayer1}, where the mapping is
explicitly given. Also, in the $SU(3)$-part an additional term is added, proportional to the
square of the second order Casimir operator of the $SU(3)$-group. This was necessary in
order to describe some non-linearities in the spectrum.

For deformed nuclei, the $SU(3)$ limit should be a good approximate symmetry. Because
in  Section \ref{examples} we discuss well deformed final nuclei and to illustrate the
application, we restrict to the $SU(3)$ symmetry limit. In spite of this limitation, in what follows
we present the general structure of the Hamiltonian for a two-cluster system.
Further, more general applications, will be presented in future.

The model Hamiltonian has the following structure:

\begin{equation}
{\bd H}  =  xy{\bd H}_{\widetilde{SU}(3)} 
+ y(1-x){\bd H}_{\widetilde{SO}(4)}+(1-y){\bd H}_{\widetilde{SO}(3)} \:\:\:,
\label{H-tot}
\end{equation}
%%%%%%
with $x$ and $y$ being mixing parameters of the dynamical symmetries
with values between 0 and 1 and 
%%%%%%
\begin{eqnarray}
{\bd H}_{\widetilde{SU}(3)} & = &
\hbar \omega \mbox{\boldmath$\widetilde{n}$}_{\pi }
+a_{4} \mathit{\widetilde{C}}_{2}\left( \widetilde{\lambda} _{C},\widetilde{\mu} _{C}\right)
\nonumber \\
&&
+(a_2-a_5\Delta
\mbox{\boldmath$\widetilde{n}$}_{\pi })\mathit{\widetilde{C}}_{2}
\left( \widetilde{\lambda} ,\widetilde{\mu} \right)  
+ t_3 \left[ \mathit{\widetilde{C}}_{2}
\left( \widetilde{\lambda} ,\widetilde{\mu} \right) \right]^2
\nonumber \\
&&
+(a_1-a_6\Delta
\mbox{\boldmath$\widetilde{n}$}_{\pi })\mathit{\widetilde{C}}_{2}
\left( \widetilde{{\bd n}}_\pi ,0 \right) 
+ t_1\mathit{\widetilde{C}}_3\left( \widetilde{\lambda} , \widetilde{\mu} \right)
\nonumber \\
&&
+\left( a_3 + a_L (-1)^{\widetilde{L}} + a_{Ln} \Delta \widetilde{{\bd n}}_\pi \right)
{\mbox{\boldmath$\widetilde{L}$}}^{2}+t_2{\mbox{\boldmath$\widetilde{K}$}}^{2}
\nonumber \\
{\bd H}_{\widetilde{SO}(4)} & = & 
a_{4} \mathit{\widetilde{C}}_{2}\left( \widetilde{\lambda} _{C},\widetilde{\mu} _{C}\right)
a_{C}{\widetilde{{\bd L}}_{C}}^{2}
+a_{R}^{\left( 1\right) }{\widetilde{{\bd L}_{R}}}%
^{2}
\nonumber \\
&&
+\left( \gamma + a_L (-1)^{\widetilde{L}} \right)
{\widetilde{{\bd L}}}^{2}
\nonumber \\
&&
+\frac{c}{4}\left[ (\mbox{\boldmath$\pi$}^{\dagger }\cdot %
\mbox{\boldmath$\pi$}^{\dagger })-(\sigma ^{\dagger })^{2}\right] \left[ (%
\mbox{\boldmath$\pi$}\cdot \mbox{\boldmath$\pi$})-(\sigma )^{2}\right] 
\nonumber \\
{\bd H}_{\widetilde{SO}(3)} & = &
\hbar \omega   \mbox{\boldmath$\widetilde{ n}$}_{\pi } 
+a_4
\mathit{\widetilde{C}}_{2}\left( \widetilde{\lambda}_{C},\widetilde{\mu}_{C}\right)  \nonumber \\
&&+a_{C}{\widetilde{{\bd L}}_{C}}^{2}+a_{R}^{\left( 1\right) }{\widetilde{{\bd L}}}%
_{R}^{2}
\nonumber \\
&&
+\left( \gamma + a_L (-1)^{\widetilde{L}} + a_{Ln} \Delta \widetilde{{\bd n}}_\pi \right)
{\mbox{\boldmath$\widetilde{L}$}}^{2}
~~~,
\label{one-d}
\end{eqnarray}
%%%%%%
where $\Delta \widetilde{{\bd n}}_\pi = \widetilde{{\bd n}}_\pi - (n_0-n_C)$, $(n_0-n_C)$
being the minimal number of quanta required by the Pauli principle
and the possible effects of the forbiddeness is taken into account by $n_C$.
The $a_{Clus}$ is the strength of the quadrupole-quadrupole
interaction, restricted to the cluster part, while $R$ and $C$
denote the contributions related to the {\it relative} and
coupled {\it cluster} part respectively, and
$\widetilde{{\bd L}}^2$ is the total angular momentum operator. 
The moment of inertia may depend on the excitation in $\tilde{n}_\pi$ 
(excited states may increase their deformation, corresponding to a larger momentum
of inertia).
The choice of (\ref{H-tot}) permits the study of phase 
transitions between, e.g., $\widetilde{SU}(3)$ and
$\widetilde{SO}(4)$ (see \cite{phase-I,phase-II} for light
nuclei).

For the case of two spherical
clusters, the second-order Casimir operator of  the pseudo-$\widetilde{SU(3)}$  is just 
$\widetilde{{\bd n}}_\pi ( \widetilde{{\bd n}}_\pi + 3)$.  Note that the information about 
the deformation of the
clusters only
enters in the $\widetilde{SU}(3)$ dynamical limit.

The first term of the $\widetilde{SU}(3)$ Hamiltonian, $\hbar\omega \mbox{\boldmath
$\widetilde{n}$}_\pi$, contains the linear invariant operator of the  $\widetilde{U}_R(3)$    
subgroup, and the $\hbar\omega$ is fixed via $(45 A^{-1/3}
- 25 A^{-2/3})$ for
light nuclei \cite{hw}, which can also be used for heavy nuclei.
For heavy nuclei $\hbar\omega = 41 A^{-\frac{1}{3}}$ is more common. The $A$ is now the mass number of the real nucleus
and not the number ${\tilde A}$ of nucleons in the normal
orbitals for the united nucleus.

The $\mbox{\boldmath $\widetilde{C}$}_2\left(SU(3)\right)$ %(SU(3))$
is the second order Casimir-invariant of the coupled
$\widetilde{SU}(3)$ group, having contributions both from the internal
cluster part and from the relative motion. It is given by:

\begin{eqnarray}
\mbox{\boldmath $C$}_2(\widetilde{SU}(3)) & = & \frac{1}{4} \mbox{\boldmath $\widetilde{Q}$}^2 +
\frac{3}{4} \mbox{\boldmath $\widetilde{L}$}^2 ,  \nonumber \\
& \rightarrow & \left(\widetilde{\lambda}^2 + \widetilde{\lambda}\widetilde{\mu} 
+ \widetilde{\mu}^2 + 3\widetilde{\lambda} + 3\widetilde{\mu}
\right) ,  \nonumber \\
\mbox{\boldmath $\widetilde{Q}$} & = & \mbox{\boldmath $\widetilde{Q}$}_C 
+ \mbox{\boldmath $\widetilde{Q}$}_R ,
\nonumber \\
\mbox{\boldmath $\widetilde{L}$} & = & \mbox{\boldmath $\widetilde{L}$}_C 
+ \mbox{\boldmath $\widetilde{L}$}_R ,
\label{su3}
\end{eqnarray}
where $\mbox{\boldmath $\widetilde{Q}$}$ and $\mbox{\boldmath $\widetilde{L}$}$ are
the quadrupole and angular momentum operator,
respectively. The relations of the quadrupole and angular momentum
operators to the $\widetilde{C}^{(1,1)}_{2m}$ generators of the $\widetilde{SU}(3)$ 
group, expressed in terms of $\widetilde{SU}(3)$-coupled $\pi$-boson creation and
annihilation operators \cite{escher}, are:

\begin{eqnarray}
\mbox{\boldmath $\widetilde{Q}$}_{k,2m} & = & \frac{1}{\sqrt{3}}
\widetilde{C}^{(1,1)}_{k2m} , \nonumber \\
\mbox{\boldmath $\widetilde{L}$}_{k1m} & = & \widetilde{C}^{(1,1)}_{k1m} , 
\nonumber \\
\mbox{\boldmath $\widetilde{C}$}^{(1,1)}_{lm} & = & \sqrt{2} \left[
\mbox{\boldmath
$\pi$}^\dagger \otimes \mbox{\boldmath $\pi$} \right]^{(1,1)}_{lm} .
\label{su3gen}
\end{eqnarray}

The operators in (\ref{su3}) and (\ref{su3gen}) 
form at the same time part of the electromagnetic transition
operators, which are defined in \cite{hess-86}.

\subsection{Spectroscopic factors}

A parametrization of the spectroscopic factor, within the SACM for light nuclei, was given in 
\cite{specfac-draayer}:

\beqa
S & = & e^{A + B n_\pi + C{\cal C}_2(\lambda_1,\mu_1) 
+ D{\cal C}_2(\lambda_2,\mu_2) 
+ E{\cal C}_2(\lambda_c,\mu_c)
+ F{\cal C}_2(\lambda , \mu ) + G{\cal C}_3(\lambda , \mu ) 
+ H\Delta n_\pi}
\nonumber \\
&&
\mid 
\langle (\lambda_1,\mu_1)\kappa_1L_1, 
(\lambda_2,\mu_2)\kappa_2L_2 \mid\mid
(\lambda_c,\mu_c)\kappa_cL_c \rangle_{\varrho_c}
\nonumber \\
&& \cdot
\langle (\lambda_c,\mu_c)\kappa_cL_c, 
(n_\pi ,0)1l \mid\mid
(\lambda ,\mu )\kappa L \rangle_1
\mid^2
~~~.
\label{specfac-light}
\eeqa
The parameters were adjusted to experimental values of 
spectroscopic factors within the p- and sd-shell,
reproducing well exact calculations within
the $SU(3)$ shell model \cite{draayer2}. For the good
agreement, the factor depending on the $SU(3)$-isoscalar factors turned out to be crucial.

For heavy nuclei spectroscopic factors are poorly or not at all known experimentally.
Therefore, we have to propose a simplified manageable ansatz, compared
to (\ref{specfac-light}), including the forbiddenness.

In what follows, we will try to get an estimate on the parameter
$B$: As argued in \cite{specfac-draayer} this term is the result of the relative part of the wave-function, which 
for zero angular momentum is proportional to 
$e^{-aR^2} \sim e^{-a\frac{\hbar}{\mu \omega} n_\pi}			 $, where
$R$ is the relative distance of the two clusters (though, an $e^{-a R}$ ansatz 
would be more appropriate, but would leave the harmonic oscillator picture)
and $a$ has units of $\rm{fm}^{-2}$. 
The $\mu$ is the reduced mass. Let us
restrict to the minimum value $n_0$ of $n_\pi$. Using the relation of 
$r_0 = \sqrt{\frac{\hbar}{\mu\omega}n_0}$ \cite{geom}, where $r_0$ is
the minimal distance between the clusters, and taking
into account that for this case $R=r_0$, we obtain
$e^{-\mid B\mid n_0}$, with 
$\mid B\mid = a\frac{\hbar}{\mu\omega}$
and $B<0$.
When the wave function is at $e^{-1}$ it gives 
$\mid B\mid =\frac{1}{n_0}$. 
For the nuclei in the sd-shell, the 
adjustment of the parameters was done for cases with $n_0=8$,
which corresponds according to the estimation to $B = -0.13$ 
approximately. This has to be compared to the value  $-0.36$ as obtained in 
\cite{specfac-draayer}, i.e., it is only a rough approximation. The most
important part of (\ref{specfac-light}) is the factor
depending on the $SU(3)$ isoscalar factors and
the influence of the exponential factor is not dominant for the relative structure of the
spectroscopic factors.
Furthermore, only {\it ratios} of spectroscopic factors are of importance,
which cancel the exponential contribution for states in the $0\hbar\omega$
shell and when the $(n_0-nc)$ is large (as it will be). Then, the corrections
for $\Delta n_\pi$ of the order of one will be negligible.
We do not see a possibility to estimate the parameter $A$ in
the exponential factor, which represents a normalization
of the spectroscopic factor.
The other terms in the exponential factor represent corrections
to the inter-cluster distance, because they correspond
to deformation effects, and the parameters in front turned out
to be consistently small. 

In light of the above estimation and discussion, 
for heavy nuclei we propose the same expression as
in (\ref{specfac-light}), but due to the not availability 
of a sufficient number of values of spectroscopic factors (or none at all)
for heavy nuclei, we propose the following simplified expression:

\beqa
S & = & e^{{\tilde A} 
+ {\tilde B} ({\tilde n}_0-{\tilde n}_C+\Delta {\tilde n}_\pi )}
\nonumber \\
&&
\mid 
\langle ({\tilde \lambda}_1,{\tilde \mu}_1)
{\tilde \kappa}_1{\tilde L}_1, 
({\tilde \lambda}_2,{\tilde \mu}_2)
{\tilde \kappa}_2{\tilde L}_2 \mid\mid
({\tilde \lambda}_c+\lambda_0,{\tilde \mu}_c+\mu_0){\tilde \kappa}_c
{\tilde L}_c \rangle_{\varrho_c}
\nonumber \\
&& \cdot
\langle ({\tilde \lambda}_c+\lambda_0,{\tilde \mu}_c+\mu_0)
{\tilde \kappa}_c{\tilde L}_c, 
({\tilde n}_\pi ,0)1{\tilde l} \mid\mid
({\tilde \lambda} ,{\tilde \mu} ){\tilde \kappa} 
{\tilde L} \rangle_1
\mid^2
~~~.
\label{specfac-heavy}
\eeqa 
The $n_\pi$ in the exponential factor was substituted
by $[(\tilde{n}_0-\tilde{n}_C)+\Delta n_\pi ]$. The $({\tilde n}_0-{\tilde n}_C)$ is the
number of relative oscillation quanta in $0\hbar\omega$ ($\tilde{n}_C$ are added to
the excitation of the clusters).
The parameter is estimated as 
${\tilde B}=-\frac{1}{{(\tilde n}_0-\tilde{n}_C)}$. Because we can not determine
the parameter ${\tilde A}$, as a consequence 
only {\it ratios} of spectroscopic factors are relevant.
An additional dependence on $\tilde{n}_C$ is contained in the product of 
reduced coupling coefficients, with the appearance of $\lambda_0$
and $\mu_0$ (see the definition in (\ref{la0mu0})). 

\subsection{Construction of the model space}
\label{sub-3.3}

In this subsection we discuss how to obtain the cluster irreps of the 
combined proton-neutron system and which further approximations have to be applied in
order that the resulting model space is not too large but still contains the
main contributions for {\it low lying states}.

In a first step, the criteria for the construction of the model space for the proton and 
neutron part are explained. The procedure is in complete analogy to the one used for
light nuclei, where now each orbital state can be occupied only by one type of nucleons 
(protons or neutrons), one with spin up and another one with spin down. Denoting by $\gamma$
= $p$ or $n$ the proton and neutron part, respectively, each cluster is in the irrep
$\left( \lambda^{\gamma}_i,\mu^{\gamma}_i\right)$ (i=1,2).
The relative motion within each subset is defined by $\left(\tilde{n}^\gamma_\pi , 0\right)$.
The ${\tilde n}^\gamma_\pi$ is the total number of relative oscillation quanta. 
The possible cluster irreps of the combined system are then, in analogy to (\ref{eq-2}), 

\beqa
&
\left( {\tilde \lambda}^\gamma_1, {\tilde \mu}^\gamma_1\right)
\otimes
\left( {\tilde \lambda}^\gamma_2, {\tilde \mu}^\gamma_2\right)
\otimes
\left( n^\gamma_\pi , 0\right)
~ \rightarrow ~
\left( {\tilde \lambda}^\gamma_C, {\tilde \mu}^\gamma_C\right)
\otimes
~~\left( n^\gamma_\pi , 0\right)
&
\nonumber \\
& \rightarrow &
\nonumber \\
& 
\left({\tilde \lambda}^\gamma ,{\tilde \mu}^\gamma \right) 
~~~.
&
\label{sub1}
\eeqa
The tildes refer to the pseudo-$SU(3)$ model.

As stated, this has to be done for each subsystem.
The question is now on how to join both systems and obtain a short list of $\widetilde{SU}(3)$
irreps? For explaining the path taken, it is useful to cast the list of irreps
in the following manner:

\beqa
&
\left\{
\begin{array}{ccc}
\left( {\tilde \lambda}^p_1, {\tilde \mu}^p_1\right) &
\left( {\tilde \lambda}^p_2, {\tilde \mu}^p_2\right) &
\left( {\tilde \lambda}^p_{C}, {\tilde \mu}^p_{C}\right) \\
\left( {\tilde \lambda}^n_1, {\tilde \mu}^n_1\right) &
\left( {\tilde \lambda}^n_2, {\tilde \mu}^n_2\right) &
\left( {\tilde \lambda}^n_{C}, {\tilde \mu}^n_{C}\right) \\
\left( {\tilde \lambda}_1, {\tilde \mu}_1\right) &
\left( {\tilde \lambda}_2, {\tilde \mu}_2\right) &
\left( {\tilde \lambda}_{C}, {\tilde \mu}_{C}\right) 
\end{array}
\right\}
&
\nonumber \\
&
\left\{
\begin{array}{ccc}
\left( {\tilde \lambda}^p_{C}, {\tilde \mu}^p_{C}\right) &
\left( {\tilde n}^p_\pi, 0\right) &
\left( {\tilde \lambda}_{p}, {\tilde \mu}_{p}\right) \\
\left( {\tilde \lambda}^n_{C}, {\tilde \mu}^n_{C}\right) &
\left( {\tilde n}^n_\pi, 0\right) &
\left( {\tilde \lambda}_{n}, {\tilde \mu}_{n}\right) \\
\left( {\tilde \lambda}_C, {\tilde \mu}_C \right) &
\left( {\tilde n}_\pi, {\tilde \mu}_\pi\right) &
\left( {\tilde \lambda}, {\tilde \mu}\right) 
\end{array}
\right\}
~~~.
&
\label{sub2}
\eeqa
The upper indices $p$ and $n$ refer to 
protons and neutrons, respectively, and the index $C$ to the cluster irrep. 
The notation in curly
brackets is intentional because it reflects the coupling of the different irreps in terms
of $9-\left(\lambda , \mu\right)$ $SU(3)$ symbols \cite{escher}. 

The first curly bracket indicates how to obtain the total cluster irrep 
$\left( {\tilde \lambda}_{C}, {\tilde \mu}_{C}\right)$ from the cluster irreps
of the proton and neutron systems. The information in  the third column is
used in the first column of the second curly bracket. 

If one takes all possibilities into account, the number of 
combinations increases astronomically, therefore, one has to find more
simplifications based on physical arguments. We suggest:

\begin{itemize}

\item When the proton fluid is combined with the neutron fluid, one can safely
assume that both fluids move coherently. When they do not, the resulting motion
corresponds to giant resonances, which are at high energies. In order to describe
them, one has to include extra interaction terms. We are not interested, however, in these
excitations and thus can assume that both fluids move coherently.

\item The assumption of a coherent motion of the proton versus the neutron fluid 
implies that when a proton irrep is coupled
to a neutron irrep, only the stretched representation is taken into account, i.e.,
$\left({\tilde \lambda}_p,{\tilde \mu}_p\right)$ $\otimes$ 
$\left({\tilde \lambda}_n,{\tilde \mu}_n\right)$ $\rightarrow$
$\left({\tilde \lambda}_p+{\tilde \lambda}_n,{\tilde \mu}_p+{\tilde \mu}_n\right)$. 
The irreps in general can refer to the individual cluster irreps, the complete cluster
irreps, the relative motion, etc. Thus, in  (\ref{sub2}) we use

\beqa
\left({\tilde \lambda}^p_1, {\tilde \mu}^p_1\right) \otimes 
\left({\tilde \lambda}^n_1, {\tilde \mu}^n_1\right)
& \rightarrow & \left( {\tilde \lambda}^p_1+{\tilde \lambda}^n_1, {\tilde \mu}^p_1+{\tilde \mu}^n_1
\right)
~=~ \left(\tilde{\lambda}_1,\tilde{\mu}_1\right)
\nonumber \\
\left({\tilde \lambda}^p_2, {\tilde \mu}^p_2\right) \otimes 
\left({\tilde \lambda}^n_2, {\tilde \mu}^n_2\right)
& \rightarrow & \left( {\tilde \lambda}^p_2+{\tilde \lambda}^n_2, {\tilde \mu}^p_2+{\tilde \mu}^n_2
\right)
~=~ \left(\tilde{\lambda}_2,\tilde{\mu}_2\right)
\nonumber \\
\left({\tilde \lambda}^p_C, {\tilde \mu}^p_C\right) \otimes 
\left({\tilde \lambda}^n_C, {\tilde \mu}^n_C\right)
& \rightarrow & \left( {\tilde \lambda}^p_C+{\tilde \lambda}^n_C, {\tilde \mu}^p_C+{\tilde \mu}^n_C
\right)
~=~ \left(\tilde{\lambda}_C,\tilde{\mu}_C\right)
\nonumber \\
\left({\tilde \lambda}_p, {\tilde \mu}_p\right) \otimes 
\left({\tilde \lambda}_n, {\tilde \mu}_n\right)
& \rightarrow & \left( {\tilde \lambda}_p+{\tilde \lambda}_n, {\tilde \mu}_p+{\tilde \mu}_n
\right)
~=~ \left(\tilde{\lambda},\tilde{\mu}\right)
\nonumber \\
\left( {\tilde n}_\pi, {\tilde \mu}_\pi\right) & \rightarrow &
\left( {\tilde n}^p_\pi + {\tilde n}^n_\pi, 0\right)
~=~ \left(\tilde{n}_\pi,0\right)
~~~,
\label{sub3-1}
\eeqa
i.e., ${\tilde \mu}_\pi = 0$.

\item When the {\it forbiddenness} is not zero, one has further to change the cluster irreps in the proton
and neutron system to the excited cluster irreps as indicated in subsection \ref{forbiddeness}.
The forbiddenness can only be calculated within each partial sector (protons or neutrons). 

\end{itemize}

With these restrictions, the number of irreps reduce considerably, but may be still 
too large for handling the calculations. A further cut-off constraint is

\begin{itemize}

\item Restrict the number of irreps in each shell $\tilde{n}_\pi$ to only the first $n_{irreps}$
with the largest eigenvalue of the second order Casimir operator of $\widetilde{SU}(3)$.
Which value to take for the cut-off value $n_{irrep}$ is a matter of choice. The justification is
that large irreps have a larger eigenvalue of the second order Casimir operator of $SU(3)$ and thus are 
lower in energy, taking into account that the coefficient should be negative as the operator
is related to the quadrupole-quadrupole interaction.

\end{itemize}

In the next section we shall illustrate the procedure for two particular cases.

\section{Applications}
\label{examples}

In this section we apply the pseudo-SACM proposed to two
sample systems. The first is
$^{236}$U $ \rightarrow$ $^{210}$Pb+$^{26}$Ne and the second one is
$^{224}$Ra $\rightarrow$ $^{210}$Pb+$^{14}$C.
For illustrative reasons, only the $\widetilde{SU}(3)$ dynamical symmetry limit will be
considered, i.e., the united nucleus must be well deformed.
A complete investigation, including studies of phase transitions between different
dynamical symmetry limits, will be presented in a future publication. 

\subsection{$^{236}$U $ \rightarrow$ $^{210}$Pb+$^{26}$Ne}

\begin{figure}
\centerline{
\includegraphics[scale=0.8]{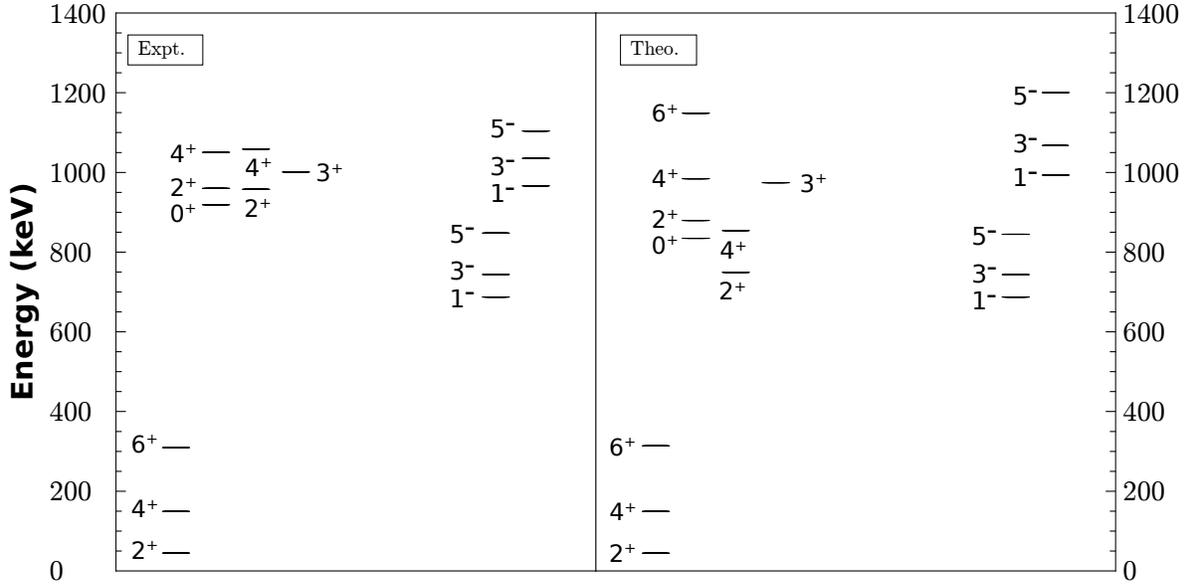} 
}
\caption{\label{Uran} 
Spectrum of $^{236}$U, described by the clusterization $^{210}$Pb+$^{26}$Ne. The theoretical spectrum
(right panel) is compared to experiment (left panel).
}
\end{figure}

The protons and neutrons are treated separately and the nucleons in each sector
are filled into the Nilsson
diagram from below, at the deformation value $\epsilon_2 =0.200$ \cite{nix-tables}.

For $^{236}$U, the united nucleus, we obtain 46 protons 
in the normal orbitals and the valence shell is
$\widetilde{\eta}_p=4$ with 6 valence protons. The ground state 
$\widetilde{SU}(3)$ irrep for the proton part
is $(\tilde{\lambda} , \tilde{\mu})_{\pi} =(18,0)_\pi$, while for the neutrons we have  82 particles 
in normal orbitals with
12 in the $\widetilde{\eta}_n=5$ valence shell, giving the ground state irrep
$(\tilde{\lambda} , \tilde{\mu})_{n} =(36,0)_n$.
These two irreps can be coupled to the total one for $^{236}$U, namely 
$(\tilde{\lambda} , \tilde{\mu}) =(54,0)$.
There are, of course, further higher lying $\widetilde{SU}(3)$ irreps in the $0\hbar\omega$ shell.
The determination of the ground state irrep is necessary for the evaluation of 
the forbiddenness (see (\ref{forbid-min})).

\begin{center}
\begin{table}[h!]
\centering
\begin{tabular}{|c|c|c|}
\hline\hline
%ind&&
$J_k^P$ & $^{236}$U: $E_{{\rm exp}}$ [MeV] & $^{224}$Ra: $E_{{\rm exp}}$ [MeV] \\
\hline
$0_1^+$ & 0.0 & 0.0 \\
$0_2^+$ & 0.919 & 0.916 \\
$0_3^+$ & - & 1.223 \\
$2_1^+$ & 0.045 & 0.084 \\
$2_2^+$ & 0.958 & 0.966 \\
$2_3^+$ & 1.094 & - \\
$2_4^+$ & 1.221 & - \\
$3_1^+$ & 1.002 & - \\
$4_1^+$ & 0.150 & 0.251 \\
$6_1^+$ & - & 0.479 \\
$1_1^-$ & 0.688 & 0.216 \\
$1_2^-$ & 0.967 & - \\
$3_1^-$ & 0.744 & 0.290 \\
\hline
$J_i^P \rightarrow J_f^P$ & $^{236}$U: $B(E2)$ [WU] & $^{224}$Ra: $B(E2)$ [WU] \\
\hline
$2_1 \rightarrow 0_1$ & 250. & 97. \\
$4_1 \rightarrow 2_1$ & 357. & 138 \\
\hline 
 \end{tabular}
\caption{
Experimental data used in the fit of the parameters of the model Hamiltonian. The second column lists
the data used for $^{236}$U and the third column for $^{224}$Ra. If no data is mentioned (dash sign), 
the value is not used in the fit.
} 
\vspace{0.2cm}
\label{fit-U-Ra}
\end{table}
\end{center}

\begin{center}
\begin{table}[h!]
\centering
\begin{tabular}{|c|c|c|}
\hline\hline
%ind&&
Parameter & $^{236}$U & $^{224}$Ra \\
\hline
$a_1$ & 0.041682 & -2.5370 \\
$a_2$ & -1.9466 & -2.5386 \\
$a_3$ & 0.014976 & 0.00029037 \\
$a_4$ & 0.46398 & -0.017288 \\
$a_5$ & -0.48563 & -0.46653\\
$a_6$ & -1.4577 & -0.46939 \\
$a_L$ & -0.0074986 & 0.012230\\
$a_{Lnp}$ & -0.016851 & 0.019355\\
$t_1$ & 0.023152 & -1.2773 \\
$t_2$ & 0.17622 & 0.42323\\
$t_3$ & 0.0016846 & 0.06194 \\
$p_{e2}$ & 1.8382 & 1.7752 \\
$e^{(2)}_{1}$ & 3.6213 & 3.9261\\
$e^{(2)}_{2}$ & 1.4332 & 1.3948 \\
\hline 
 \end{tabular}
\caption{
List of parameter values obtained from the fit, for $^{236}$U in the second column and
for $^{224}$Ra in the third column. Also listed are the effective charges used in the fit.
The effective charges are defined in \cite{hess-86}, where
$e^{(2)}_{k}$ is the effective charge in the quadrupole transition operator for 
cluster no. $k=1,2$, estimated geometrically, and $p_{e2}$ is a factor describing
the deviation from the geometrical estimate. For details, please consult \cite{hess-86}. 
} 
\vspace{0.2cm}
\label{para-U-Ra}
\end{table}
\end{center}

\begin{center}
\begin{table}[h!]
\centering
\begin{tabular}{|c|c|c|c|c|}
\hline\hline
%ind&&
$J_i^{P_i} \rightarrow J_f^{P_f}$ & $^{236}$U (th) & $^{236}$U (exp) & $^{224}$Ra (th) & 
$^{224}$Ra (exp) \\
\hline
$2_1^+ \rightarrow 0_1^+$ & 250. & 250. & 97.3 & 97.0 \\
$2_1^+ \rightarrow 2_2^+$ & 0.565 & - & 5.81 & - \\
$4_1^+ \rightarrow 2_1^+$ & 356. & 357.  & 138. &  138. \\
$2_1^+ \rightarrow 3_1^+$ & 0.831 & - & 32.0 & - \\
$3_1^+ \rightarrow 4_1^+$ & 0.353 & - & 8.17 & - \\
$1_1^- \rightarrow 2_1^-$ & 545. & - & 42.7 & - \\
$3_1^- \rightarrow 2_1^-$ & 310. & - & 28.1 & - \\
\hline 
 \end{tabular}
\caption{
Theoretical calculated $B(E2,J_i^{P_i} \rightarrow J_f^{P_f})$ transition values in Weisskopf
units, compared to the experimental values, if available. The first column indicates the transition,
the second column the theoretical values for $^{236}$U and the third column the corresponding
experimental values, if available. The corresponding data for $^{224}$Ra are listed in the 
last two columns. 
} 
\vspace{0.2cm}
\label{Trans-U-Ra}
\end{table}
\end{center}

\begin{center}
\begin{table}[h!]
\centering
\begin{tabular}{|c|c|c|}
\hline\hline
%ind&&
$J_k^P$ & $^{236}$U (th) & $^{224}$Ra (th) \\
\hline
$0_1^+$ & 0.0015 & 0.0069 \\
$0_2^+$ & 0.0015 & 0.0069 \\
$2_1^+$ & 0.0015 & 0.0063 \\
$2_2^+$ & 0.0 & 0.0063 \\
$4_1^+$ & 0.014 & 0.0048 \\
$4_2^+$ & 0.0  & 0.0048 \\
$1_1^-$ & 0.0  & 0.0065 \\
$2_1^-$ & 0.0  & 0.0058 \\
$3_1^-$ & 0.0  & 0.0055 \\
\hline
\hline 
 \end{tabular}
\caption{
Some spectroscopic factors, divided by $e^{\widetilde{A}}$, of low lying states. 
In the first column the state considered is listed.
The values of the spectroscopic factor for $^{236}$U and $^{224}$Ra
are in the second and third column, respectively.
} 
\vspace{0.2cm}
\label{Spec-U-Ra}
\end{table}
\end{center}

These considerations have to be repeated for the two clusters involved. The largest cluster is
$^{210}$Pb. Filling the protons into the Nilsson diagram, at the same deformation as for the
united nucleus, we obtain 40 protons in normal orbitals, where the valence shell is
${\widetilde \eta}=3$ and closed, thus the corresponding irrep is $(0,0)_{p}$. For the neutrons one has
72 in normal orbitals with 2 neutrons in the ${\widetilde N}=5$ pseudo-shell. The corresponding
irrep is $(10,0)_{n}$.

The light cluster $^{26}$Ne is {\it put on top} of the heavy cluster. We count 6 protons and 10 neutrons
in normal orbitals, which gives $(0,2)_{p}$ and $(4,0)_{n}$.

The minimal number of quanta which have to be added in the proton part is 20, corresponding 
to a $(20,0)_{p R}$ 
irrep in the relative part. For the neutron part, this number is 40, i.e.,
an irrep $(40,0)_{n R}$.

In the next step, the proton parts of the clusters are coupled with the relative part of the protons. The same is done for the neutrons. For the proton part, the product
$(0,0)_p \otimes (0,2)_p \otimes (20,0)_{p R}$ contains the proton irrep $(18,0)_p$
of the united nucleus, thus, the forbiddenness for the proton part is zero. 
The situation is different for the neutron part: The product 
$(10,0)_n \otimes (4,0)_n \otimes (40,0)_{n R}$
{\it does not contain} (36,0), which is the irrep in the united nucleus.
This indicates that one has to excite the clusters and the forbiddenness is different from zero.
Using the formula (\ref{forbid-min}) we obtain a forbiddennes of $n_C=2$. The excitation of the clusters
is achieved, changing the irrep of $^{26}$Ne from $(4,0)_n$ to $(6,0)_n$. 
The relative part is now reduced by two 
quanta, leaving $(38,0)_{nR}$. With this change, the product
$(10,0)_n \otimes (6,0)_n \otimes (38,0)_{n R}$ 
now contains the dominant irrep for neutrons in $^{236}$U.

Using the Hamiltonian (\ref{H-tot}) in the $SU(3)$-dynamical limit, the coefficients are adjusted to 
the experimental data, listed in Table \ref{fit-U-Ra} in the second column. 
The optimal parameters obtained are listed in
Table \ref{para-U-Ra}, second column. 
With these parameters, the spectrum calculated is depicted in Figure \ref{Uran}.
The calculated B(E2)-transition values are listed in Table \ref{Trans-U-Ra}, second (theory)
and third (experiment) column. 

As can be noted, the agreement to experiment is satisfactory and shows the effectiveness of the
pseudo-SACM to describe the collective structure of heavy nuclei.

Next, we calculated some spectroscopic factors, listed in Table \ref{Spec-U-Ra},
second column. The Equation (\ref{specfac-heavy}) was used with the approximation of the
parameter $\widetilde{B}$ as $\left(-\frac{1}{{\tilde n}_0-n_C}\right)$. The total number of relative oscillation
quanta for the system under study is ${\tilde n}_0=60$, thus $\widetilde{B} \approx 0.0172$ and the exponential
factor in (\ref{specfac-heavy}) acquires the form 
$e^{\widetilde{A}-0.0172({\tilde n}_0-n_c+\Delta{\tilde n}_\pi )}$
$\approx$ $(0.983)^{({\tilde n}_0-n_c+\Delta{\tilde n}_\pi )}e^{\widetilde{A}}$. 
The factor $e^{\widetilde{A}}$ is unknown and thus
in Table \ref{Spec-U-Ra}, the ratios of the spectroscopic factors are more trust-worthy. 
As can be observed,
the spectroscopic factors to $\Delta {\tilde n}_\pi = 1$ are suppressed, 
where we listed values smaller than $10^{-5}$ as zero.

\subsection{$^{224}$Ra $\rightarrow$ $^{210}$Pb+$^{14}$C}

\begin{figure}
\centerline{
\includegraphics[scale=0.8]{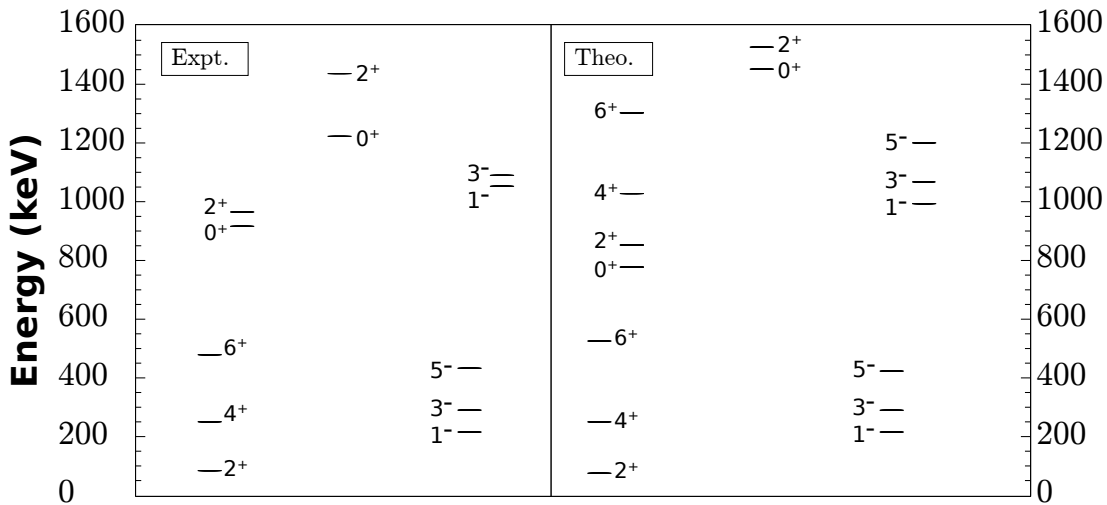} 
}
\caption{\label{Ra} 
Spectrum of $^{224}$Ra, described by the clusterization $^{210}$Pb+$^{14}$C. The theoretical spectrum
(right panel) is compared to experiment (left panel).
}
\end{figure}

As in the former section, the protons and neutrons are treated separately, 
where the nucleons are filled into the Nilsson
diagram from below, at the deformation value $\epsilon_2 =0.150$ \cite{nix-tables}.

For $^{224}$Ra, the united nucleus, we obtain 44 protons 
in the normal orbitals and the valence shell is
${\widetilde \eta}_p=4$ with 4 valence protons. The $\widetilde{SU}(3)$ irrep
is $(\tilde{\lambda} , \tilde{\mu})_{p} =(12,2)_p$, while for the neutrons we have   
78 normal particles with 8 in the ${\widetilde \eta}_n=5$ valence shell, 
giving $(\tilde{\lambda} , \tilde{\mu})_{n} =(26,4)_n$.
These two irreps can be coupled to the total one for $^{224}$Ra, namely 
$(\tilde{\lambda} , \tilde{\mu}) =(38,6)$.

These considerations have to be repeated for the two clusters involved. The largest cluster is
$^{210}$Pb, with the same numbers as in the former sub-section. 

The light cluster $^{14}$C is added on top of the heavy cluster. We count 4 protons and 6 neutrons
in normal orbitals, which gives $(2,0)_{p}$ and $(0,2)_{n}$.

The minimal number of quanta which have to be added in the proton part is 14, 
corresponding to a ($14,0)_{p R}$ 
irrep in the relative part. For the neutron part, this number is 26, i.e.,
an irrep $(26,0)_{n R}$.

For the united nucleus, 
the proton part is coupled separately to the neutron part. For the proton part, the product
$(0,0)_p \otimes (2,0)_p \otimes (14,0)_{p R}$ contains the proton irrep $(12,2)_p$
of the united nucleus, thus, the forbiddenness for the proton part is zero. 
For the neutron part the product $(10,0)_n \otimes (0,2)_n \otimes (26,0)_{n R}$
{\it does now contain} $(26,4)_n$, which is the irrep in the united nucleus.
This shows that the forbiddenness in this case is zero.

Using the Hamiltonian (\ref{H-tot}) in the $SU(3)$-dynamical limit, the coefficients are adjusted to 
the experimental data, listed in Table \ref{fit-U-Ra}, third column. The optimal parameters obtained are listed in
Table \ref{para-U-Ra}, third column. 
With these parameters, the spectrum calculated is depicted in Figure \ref{Ra}.
The calculated B(E2)-transition values are listed in Table \ref{Trans-U-Ra}, 
fourth (theory) and fifth (experiment) column. 

As can be noted, the agreement to experiment is satisfactory and shows 
also in this example the effectiveness of the
pseudo-SACM to describe the collective structure of heavy nuclei.

Next, we calculated some spectroscopic factors for $^{224}$Ra, listed in Table \ref{Spec-U-Ra},
third column. The total number of relative oscillation
quanta for the system under study is ${\tilde n}_0=40$ ($n_C=0$), 
thus $\widetilde{B} \approx 0.025$ and the exponential
factor in (\ref{specfac-heavy}) acquires the form 
$e^{\widetilde{A}-0.025({\tilde n}_0-n_c+\Delta{\tilde n}_\pi )}$
$\approx$ $(0.975)^{({\tilde n}_0-n_c+\Delta{\tilde n}_\pi )}e^{\widetilde{A}}$. 
The factor $e^{\widetilde{A}}$ is unknown and thus
in Table \ref{Spec-U-Ra}, the spectroscopic factors are divided by $e^A$. 
As can be observed, in contrast to $^{236}$U, now 
the spectroscopic factors to $\Delta {\tilde n}_\pi = 1$ are not suppressed
and are of the same order as those to $\Delta {\tilde n}_\pi = 0$.

\section{Conclusions}
\label{conclusions}

We have presented an extension of the {\it Semimicroscopic Algebraic Cluster Model} (SACM),
for light nuclei, to the {\it pseudo-SACM}, for heavy nuclei. 
Though, there exist former attempts to extend the
SACM to heavy nuclei, we found it necessary to construct a model, which enables us to 
circumvent some problems of the former approaches and to deliver a more consistent
procedure. 

In order to extend the SACM to heavy nuclei, several basic assumptions, philosophies and 
procedures had to be explained, as the concept of {\it forbiddenness}, the use
of the same deformation and $\hbar\omega$ for the clusters and the united nucleus.
Further constraints, as the assumption that the proton and neutron part couple only linearly,
were implemented.

For illustrative reasons, the applications were restricted to the dynamical 
$\widetilde{SU}(3)$ limit. As examples, we considered 
$^{236}$U $ \rightarrow$ $^{210}$Pb+$^{26}$Ne and
$^{224}$Ra $\rightarrow$ $^{210}$Pb+$^{14}$C. We demonstrated that the model is
able to describe the spectrum and electromagnetic transition probabilities. 
Spectroscopic factors were also calculated, without further fitting and they can be
considered as a prediction of the model. 

The restriction to the $\widetilde{SU}(3)$ dynamical symmetry limit has to be relaxed
in future applications, including the other 
dynamical symmetries inherit in the Hamiltonian (\ref{H-tot}). 
Also the study of phase transitions are of
interest, requiring the use of the geometrical mapping \cite{geom} of the SACM.

In future, we will also study applications to other heavy systems, especially those
where the two clusters are nearly equal. In this case, it is not clear which cluster
we have to select first in order to start the filling of the $\widetilde{SU}(3)$
Nilsson levels. The microscopic model space probably will differ slightly when one
or the other path is taken, not changing the overall structure.

\section*{Acknowledgments}
We acknowledge financial support form DGAPA-PAPIIT (IN100315) and to
CONACyT (project number 251817).
Very useful discussions with J. Cseh (ATOMKI, Hungary) are acknowledged.

\end{document}